\documentclass[letterpaper,twocolumn]{jpsj3}
\usepackage{txfonts}
\usepackage{bm}
\usepackage{color}
\newcommand{\mibs}[1]{\mbox{\small\boldmath $#1$}}

\title{A Series Expansion Study for Large Negative Quantum Renormalization of Magnon Spectra in the $S=1/2$ Kagome-Lattice Heisenberg Antiferromagnet Cs$_{2}$Cu$_{3}$SnF$_{12}$}

\author{Singo Kogure$^1$, Masashi Takeda$^1$, Katsuhiro Morita$^1$, Yoshiyuki Fukumoto$^1$\thanks{yfuku@rs.tus.ac.jp}, Mutsuki Saito$^2$, and Hidekazu Tanaka$^3$}
\inst{$^1$Department of Physics, Faculty of Science and Technology, Tokyo University of Science, Noda, Chiba 278-8510, Japan\\
$^2$Department of Physics, Tokyo Institute of Technology, Meguro-ku, Tokyo 152-8551, Japan\\
$^3$Innovator and Inventor Development Platform, Tokyo Institute of Technology, Nagatsuda, Midori-ku, Yokohama 226-8502, Japan} 

\abst{
The series expansion method is used to study magnon spectra of the kagome system with nearest-neighbor exchange interaction $J$ 
and out-of-plane Dzyaloshinskii-Moriya (DM) interaction $D^{\parallel}$, which is a minimal model for Cs$_{2}$Cu$_{3}$SnF$_{12}$.
Compared to the magnon spectra by the linear spin wave (LSW) theory, we find that dispersions at high energy part suffer downward deformation, which is similar to the triangle lattice case,
in addition to the reduction of the energy scale of about 40\% as pointed out in a neutron-scattering study by Ono {\it et al.}
Using a reliable estimation $J=20.7$ meV in a previous study on the magnetic susceptibility of Cs$_{2}$Cu$_{3}$SnF$_{12}$, 
we use $D^{\parallel}$ as the fitting parameter to reproduce the experimental magnon spectra and obtain $D^{\parallel}=0.12J$. 
We also report that a roton-like minimum occurs at the M point and a maximum at points somewhat away from the $\Gamma$ point.
Compression of the LSW magnon band is commonly seen in the kagome and triangular-lattice systems, which may be viewed as being pushed from above by the spinon continuum.
}

\begin{document}
\maketitle


Two-dimensional spin-1/2 frustrated spin systems have received particular attention, since P.~W.~Anderson predicted in 1973 that the resonating valence bond (RVB) state would be realized.\cite{Anderson1973}
Elementary excitation from the RVB state is a spinon, and a triplet excitation consists of two spinons.
In recent years, spin-1/2 frustrated spin systems {\it with magnetic orderings} have attracted renewed interests
because neutron scattering experiments on a triangular-lattice system Ba$_3$CoSb$_2$O$_9$\cite{Ito2017} and a kagome-lattice system Cs$_2$Cu$_3$SnF$_{12}$,\cite{Ono2014,Saito2022,Matan2022} in which the $\mib{q}=0$ type ordering with positive chirarity is stabilized by Dzyaloshinskii-Moriya (DM) interactions,\cite{Elhajal2002}  have revealed that the dynamic structure factors consist of not only collective magnon excitations but also spinon continuums. 

As for the triangular-lattice system, Zhang and Li used the fermionic spinon operator to calculate dynamic structure factors within the RPA approximation.\cite{Zhang2020}
They argued that the structure of the spinon continuum is closely related to the roton minimum of the magnon spectrum, and also found an amplitude mode composed of two bound spinons that cannot be captured by the spin wave approximation. 

As for the kagome-lattice system, the magnon dispersions obtained from a neutron scattering experiment on Cs$_2$Cu$_3$SnF$_{12}$ were analyzed based on the linear spin wave (LSW) approximation, and it was reported that the spin-wave interactions led to a large negative-renormalization where the exchange parameter was found to be about 40\% smaller than a reliable value, $J=20.7$ meV, determined from magnetic susceptibility measurements.\cite{Ono2014}
Also, a recent neutron scattering experiment has observed a fourth collective excitation that is thought to be an amplitude mode, in addition to the spinon continuum.\cite{Saito2022,Matan2022}
The fourth mode was also found theoretically in a recent variational Monte Carlo study.\cite{Ferrari2023}
In this article, we report our recent results of an Ising expansion (IEP) study on the large negative-renormalization observed in Cs$_2$Cu$_3$SnF$_{12}$. 

\begin{figure}[b]
\centering
\includegraphics[clip,width=8.cm]{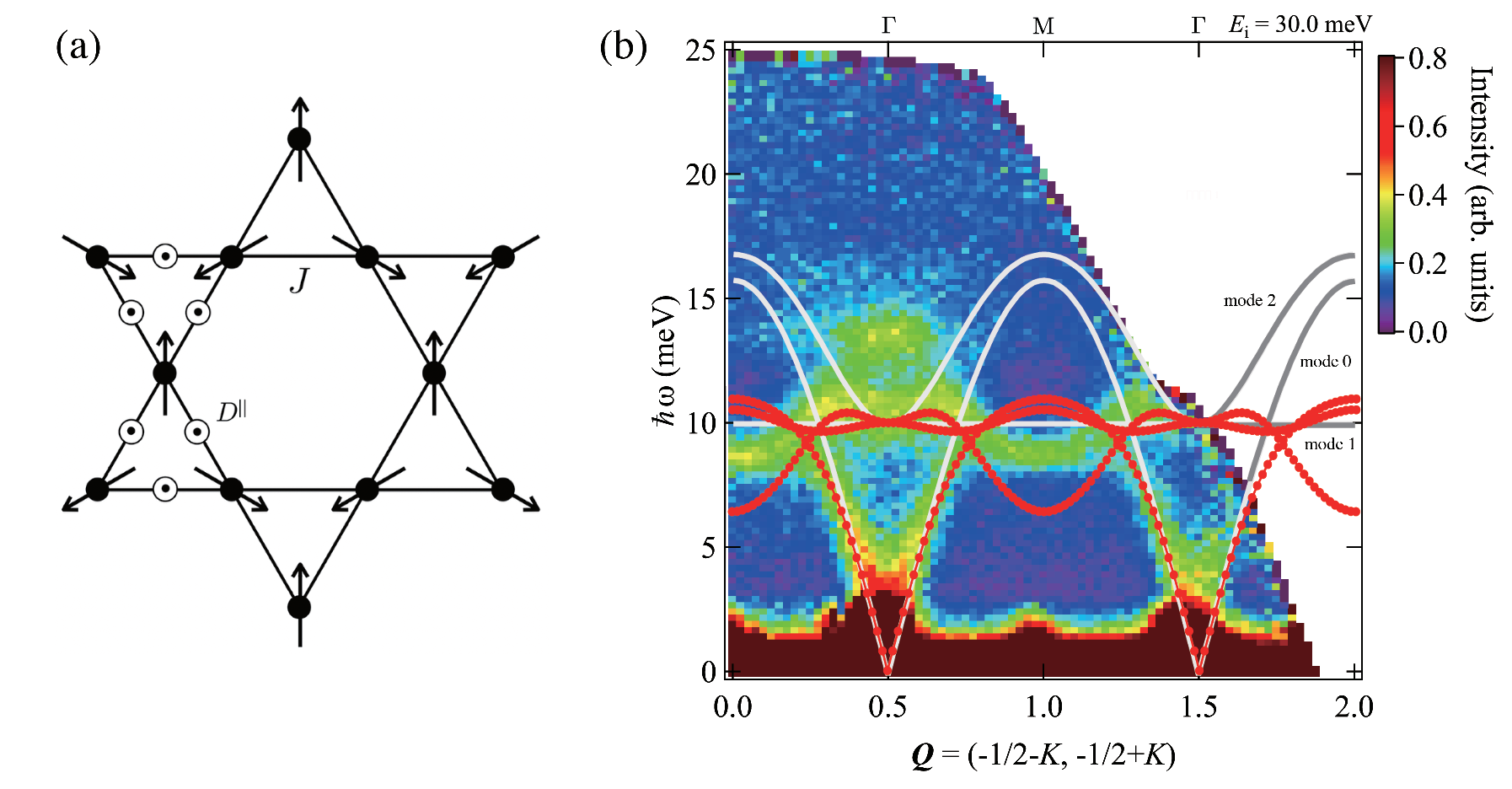}
\caption{
(Color online) (a) The arrows represent the direction of magnetic moment in the $\mib{q}=0$ ordering state,
and the unit vector $\mib{n}$ in $\mib{D}=D^{\parallel}\mib{n}$ has a direction toward the front from the page.
(b) Comparison of the results of neutron scattering experiment (color map),\cite{Saito2022} LSW (gray line), and ninth-order IEP (red line) with $t/J=1$.
As for model parameters, we have used $J=12.8$meV, $D^{\parallel}/J=0.18$ for LSW, and $J=20.7$meV, $D^{\parallel}/J=0.12$ for IEP.
}
\label{fig:dminteraction_direction}
\end{figure}

In order to stabilize the $\mib{q}=0$ ordering shown in Fig.~\ref{fig:dminteraction_direction}(a), we incorporate the DM interaction with the $D$ vector of $\mib{D}=D^{\parallel}\mib{n}$, where $\mib{n}$ represents the normal vector of the plane containing the magnetic moments,  and write a minimal Hamiltonian for Cs$_{2}$Cu$_{3}$SnF$_{12}$ as
\begin{align}
   H=J\sum_{\langle i,j\rangle}\bm{S}_{i}\cdot\bm{S}_{j}+\sum_{\langle i,j\rangle} \bm{D}\cdot(\bm{S}_{i} \times \bm{S}_{j}), 
   \label{eq:1}
\end{align}
where $\langle i,j\rangle$ and $J$ denote nearest neighbors and exchange coupling respectively, and $\boldsymbol{S_{\textit{i}}}$ is a spin-1/2 operator at site $i$.
Because the second term contains a cross product, we order the three sites in each triangle clockwise and regard site $i$ ($j$) as the prior (subsequent) site to uniquely define the Hamiltonian.
When $D^{\parallel}>0$, the $\bm{q}=0$ magnetic ordering with positive chirality shown in Fig.~\ref{fig:dminteraction_direction}(a) is stabilized.

Figure~\ref{fig:dminteraction_direction}(b) compares the magnon spectrum of (\ref{eq:1}) calculated by LSW with the dynamic structure factor obtained by neutron scattering in Ref.~4.
The path on the wavenumber plane is from the $\Gamma$ point to the M point.
The color map represents the experimental dynamic structure factor.
The gray lines represent the LSW results, which are composed of three modes.
The first is acoustic mode, which we call mode 0.
Mode 0 is gapless at the $\Gamma$ point and has an excitation energy of 15 meV at the M point.
There are two optical modes.
One is completely flat, which we call mode 1.
Mode 1 has an excitation energy of 10~meV, which becomes a zero energy mode at $D^{\parallel}\rightarrow 0$.
Another optical mode, which we call mode 2, degenerates to mode 1 at the $\Gamma$ point and has an excitation energy of 16 meV at the M point.
Interpreting the experimental dynamic structure factor in terms of the LSW results, mode 1 was observed over the entire wavenumber range of measurements,\cite{Ono2014,Saito2022,Matan2022} 
and mode 0 and mode 2 were observed only around the $\Gamma$ point.
In the color map, we notice the appearance of the fourth mode around 14 meV near the $\Gamma$ point, which has been presumed to be an amplitude mode that cannot be captured by LSW.\cite{Saito2022}
Finally, and most importantly, $J=12.8$meV and $D^{\parallel}/J=0.18$ have been used in the calculation of the LSW, but the exchange interaction $J=12.8$meV is much smaller than $J=20.7$meV obtained from high temperature susceptibility analysis.\cite{Ono2014}
The former is about 40\% smaller than the latter, and Ono {\it et al.} pointed out that this ``large negative-renormalization'' occurs due to spin-wave interactions.\cite{Ono2014}
On the other hand, by comparing the magnon spectra of the triangular lattice system obtained by LSW and IEP, it is known that the flattening of high-energy part occurs in the triangular lattice system and low-energy part is not affected very much by spin-wave interactions (see Fig.~5 in Ref.~10).
In the dynamic structure factor of Cs$_{2}$Cu$_{3}$SnF$_{12}$, only one mode has intensity around the M point.
Although the LSW predicts excitation energies of $\sim 15$~meV for mode 0 and mode 2 around the M point, but it will be interesting to see what happens with IEP.
The results of the IEP are shown by the red lines in Fig.~\ref{fig:dminteraction_direction}(b), and our conclusion is that flattening occurs in mode 0 and a roton-like minimum occurs in mode 2.

Here it is instructive to see the result of $S=5/2$ kagome antiferromagnet KFe$_3$(OH)$_6$(SO$_4$)$_2$.
Matan {\it et al.} measured the magnon spectra using inelastic neutron scattering.\cite{Matan2006}
Mode 0 and/or mode 2 were observed even around the M point, and the measured magnon spectra were well fitted by the LSW (see Fig.~3 in Ref.~11).
The exchange parameter was estimated as $J=3.18$~meV through the fitting.
On the other hand, a susceptibility measurement gave a somewhat larger value $J=3.9(2)$~meV.\cite{Grohol2005}
The former is about 17\% smaller than the latter.
Thus the effect of renormalization for $S=5/2$ is much smaller than for $S=1/2$.

Now we start describing our IEP calculations.
It is convenient to introduce local coordinates for each site, in which the $z$ axis is chosen to be parallel to the direction of the magnetic moment. 
Then, the Hamiltonian in eq.~(\ref{eq:1}) can be written as
\begin{align}
   H&&\hspace{-12mm}=\sum_{\langle i,j\rangle} \Big\{-E_z S_i^z S_j^z-E_x S_i^x S_j^x -E_y S_i^y S_j^y 
   \nonumber\\
   &&-d_y(S_i^x S_j^z-S_i^z S_j^x)\Big\},
\label{eq:3}
\end{align}
where 
$E_x=E_z=\frac{J+\sqrt{3}D^{\parallel}}{2}$, 
$E_y=-J$, 
$d_y=\frac{-\sqrt{3}J+D^{\parallel}}{2}$.
Introducing a perturbation parameter $\lambda$ and a local field term, we extend (\ref{eq:3}) as 
\begin{align}
   H_{\lambda}=H_0+\lambda H_1,
\label{eq:4}
\end{align}
where
\begin{align}
   H_0=-E_z\sum_{\langle i,j\rangle}  S_i^z S_j^z-t\sum_i\left(S_i^z-\frac{1}{2}\right)
\label{eq:5}
\end{align}
and
\begin{align}
   H_1=&\sum_{\langle i,j\rangle} \Big\{-E_x S_i^x S_j^x -E_y S_i^y S_j^y -d_y(S_i^x S_j^z-S_i^z S_j^x)\Big\} 
   \nonumber\\
   &+t\sum_i\left(S_i^z-\frac{1}{2}\right).
\label{eq:6}
\end{align}
We have included the local field term of strength $t$ into $H_0$ in eq.~(\ref{eq:5}) and that with the opposite sign into $H_1$ in eq.~(\ref{eq:6}), so as to cancel out with each other at $\lambda\rightarrow 1$.
These terms are important for improving the convergence of the series.
It was introduced originally in the series expansion study of the triangular lattice,\cite{Zheng2006} where calculations with $t=J,\;2J$ were carried out.


Using $H_{\lambda}$, we make multi-block diagonalizations to obtain the series expansion of irreducible matrix elements of the effective Hamiltonian in the single-flip sector in terms of $H_0$.\cite{Zheng2006}
For each wave vector $\mib{k}=(k_x,k_y)$, the irreducible matrix elements can be written as
\begin{align}
   \Delta_{\nu,\nu'}(k_x,k_y)=\sum_{r=0}^{\infty}\lambda^r\sum_{m,n}c_{\nu,\nu'}(r,m,n)f(m,n;k_x,k_y),
\label{eq:7}
\end{align}
where $\nu$ ($\nu'$) is the index for the three spins in an unit cell of the kagome lattice, 
$c_{\nu,\nu'}(r,m,n)$ represents the $r$th-order term of the matrix element by which the flipped spin at $\nu'$th site on the unit cell at the origin moves to $\nu$th site on the unit cell $(m,n)$, and
\begin{align}
   f(m,n;k_x,k_y)=&\frac{1}{3}\Bigg
   \{\cos(mk_x)\cos(\sqrt{3}nk_y)
   \nonumber\\
   &+\cos\left(\frac{m+3n}{2}k_x\right)\cos\left[\frac{\sqrt{3}(m-n)}{2}k_x\right]
   \nonumber\\
   &+\cos\left(\frac{m-3n}{2}k_x\right)\cos\left[\frac{\sqrt{3}(m+n)}{2}k_x\right]
   \Bigg\}.
\end{align}
 We calculated all clusters up to 9 links and 10 sites, and obtained all matrix elements up to $r=9$.
By diagonalizing the $3\times 3$ matrix whose elements are given by (\ref{eq:7}), we obtain the series up to the ninth order of the magnon excitation energy $\epsilon_{\mibs{k},\alpha}$,
where $\alpha=0,\;1,\;2$ identifies modes 0, 1, 2.

\begin{figure}[t]
\centering
\hspace{-.5cm}
\includegraphics[clip,width=6.5cm]{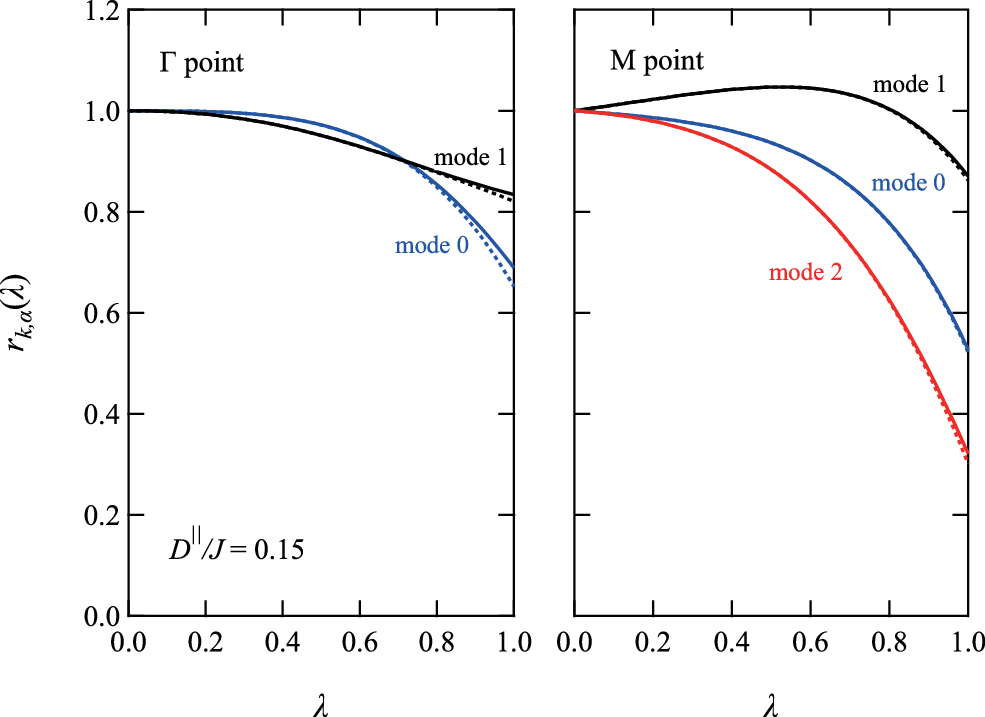}
\caption{(Color online) (a) Renormalization factors $r_{k,\alpha}(\lambda)$ by the IEP with $t/J=1$.
The solid and dashed lines, respectively, represent the ninth and eighth order results.
Note that mode 2 degenerates to mode 1 at the $\Gamma$ point.}
\label{fig:lambda_dependence}
\end{figure}

We are interested in the renormalization factor for the LSW magnon excitation energy due to spin-wave interactions.
Thus, we write the magnon excitation energy $\epsilon_{k,\alpha}$ as
\begin{equation}
    \epsilon_{k,\alpha}=r_{k,\alpha}\times\epsilon_{k,\alpha}^{\rm{LSW}},
    \label{eq:9}
\end{equation}
where $r_{k,\alpha}$ is the renormalization factor.
The LSW magnon excitation energy, $\epsilon_{k,\alpha}^{\rm{LSW}}(\lambda)$, is obtained by treating (\ref{eq:4}) with the Holstein-Primakoff transformation, and it is converted to the power series in $\lambda$ up to the $n$th order term,
which is represented by $\epsilon_{k,\alpha}^{(n)\rm{LSW}}(\lambda)$.
On the other hand, we set $\epsilon_{k,\alpha}^{(n)\rm{IEP}}(\lambda)$ the $n$th order series of the magnon excitation energy obtained by the IEP.
From the ratio $\epsilon_{k,\alpha}^{(n)\rm{IEP}}(\lambda)/\epsilon_{k,\alpha}^{(n)\rm{LSW}}(\lambda)$, we can obtain $n$th order series, $r_{k,\alpha}^{(n)}(\lambda)$, for the renormalization factor.
Although for $\lambda\rightarrow 1$ there appears a square-root singularity in $\epsilon_{k,\alpha}^{\rm{LSW}}(\lambda)$ of mode 0 at the $\Gamma$ point, we assume the renormalization factor has no such singularity.
We calculate $r_{k,\alpha}(\lambda)$ by the IEP up to the ninth order, and evaluate value of $r_{k,\alpha}$ in eq.~(\ref{eq:9}) by naive summation of the obtained series.
In Fig.~\ref{fig:lambda_dependence} we show the eighth and ninth order results of $r_{k,\alpha}(\lambda)$ at the $\Gamma$ and M points as a function of $\lambda$, and find a good convergence.
In particular, for the M point, the renormalization factor for mode 2 amounts to $0.3$ at $\lambda\rightarrow 1$, which demonstrates that large negative renormalization occurs in the kagome system.

\begin{figure}[t]
\centering
\includegraphics[clip,width=5.3cm]{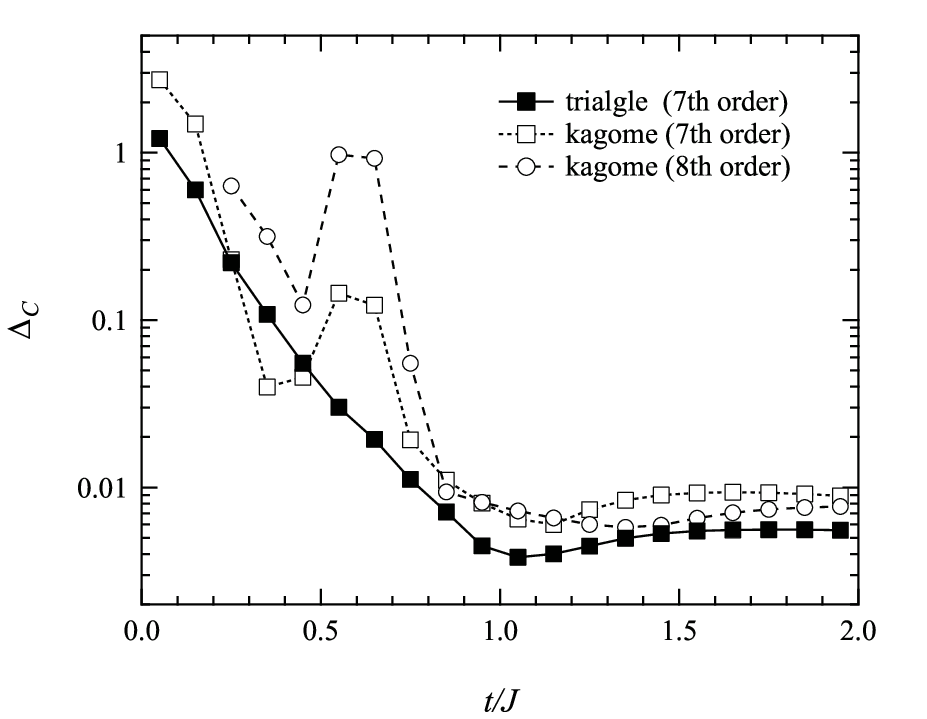}
\caption{$\Delta_{C}$ as a function of $t/J$.
For the kagome system, $D^{\parallel}/J=0.15$ is chosen.
}
\label{fig:t_depend_kagome}
\end{figure}

We here discuss the choice of value of $t$.
The $t$ term was introduced in the study of triangular lattice systems.
If the $t$ term is not introduced, a region where the magnon energy is negative will appear in the wavenumber space, and unphysical results will be obtained.
The introduction of the $t$ term improves this point.
However, the result becomes dependent on the value of $t$.
It is desirable to select a value for $t$ that minimizes this dependence as much as possible.
We introduce the following quantities in order to understand the sensitivity of the calculation results to $t$:
\begin{equation}
  \Delta_{C}(t)=\frac{\sum_{\alpha}\int_C dk\left|\epsilon_{k,\alpha}^{t+\delta/2}-\epsilon_{k,\alpha}^{t-\delta/2}\right|}{J\int_C dk}
\end{equation}
with $\delta=0.1J$. 
Here, $C$ is a path in wavenumber space, and for the triangular lattice system, consider the one in Figure 5 of Ref.~10, and for the kagome lattice system, consider the one in Fig.~1(b).
Also, $\alpha$ is a mode identifier, and there is one type of $\alpha$ for the triangular lattice system, and three types of $\alpha$ for the kagome lattice system.
Calculated results are shown in Fig.~\ref{fig:t_depend_kagome}.
The filled squares represent $\Delta_{C}(t)$ for the triangle lattice.
When $t>0$, the results change significantly at first, and the region of negative magnon energy decreases.
When $t$ increases to some extent, the region of negative magnon energy disappears and the result does not depend on $t$ very much.
$\Delta_C$ is minimized around $t/J=1$.
As $t$ is further increased, the resulting $t$ dependence increases slowly, because it becomes harder to correctly handle the local field term included in the perturbation.
For a triangular lattice, $t/J=1$ is a good choice from the viewpoint of reducing the sensitivity.
The open symbols in Fig.~\ref{fig:t_depend_kagome} represent $\Delta_{C}(t)$ for the kagome lattice.
In the kagome lattice, an increase is seen at $t/J\sim 0.6$. However, when $t/J>\sim 0.8$, $\Delta_C<10^{-2}$.
The minimum value of the susceptibility is slightly larger than that of the triangular lattice.
The minimum value of susceptibility is obtained slightly beyond $t/J=1$, but in this study, $t/J=1$ is used in calculations for kagome as well.

\begin{figure}[b]
\centering
\includegraphics[clip,width=7.5cm]{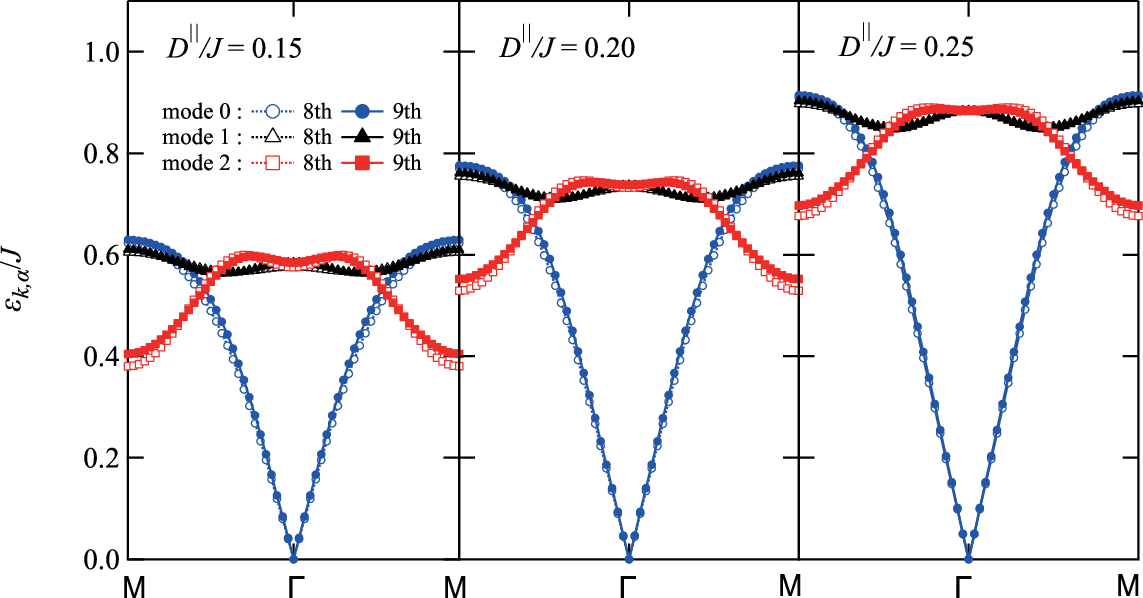}
\caption{(Color online) Calculated magnon dispersion relations by the IEP with $t/J=1$ at 8th and 9th orders.
}
\label{fig:disp_se}
\end{figure}

\begin{figure}[t]
\centering
\includegraphics[clip,width=7.5cm]{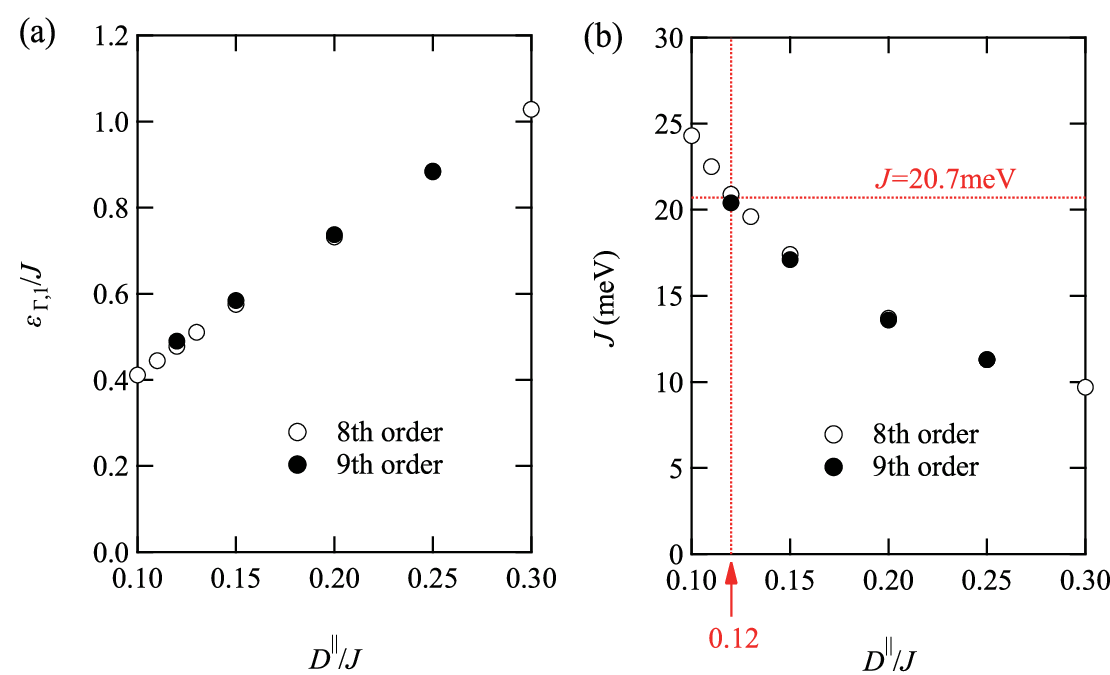}
\caption{(Color online) (a) Excitation energy at the $\Gamma$ point, $\epsilon_{\Gamma,1}/J\;(=\epsilon_{\Gamma,2}/J)$, and
(b) value of exchange parameter $J$ corresponding to $\epsilon_{\Gamma,1}=10$ meV as functions of $D^{\parallel}/J$
by the IEP with $t/J=1$ at 8th and 9th orders.
}
\label{fig:J_estimation}
\end{figure}

Figure~\ref{fig:disp_se} shows the magnon dispersion relation up to the ninth-order IEP for $D^{\parallel}/J=0.15$, 0.20, and 0.25.
The last one is a theoretical model for YCu$_{3}$(OH)$_{6}$Cl$_{3}$.\cite{Arh2020}
As $D^{\parallel}/J$ increases, the bandwidth increases, but the shape of dispersion curves is almost the same.
To investigate the dependence of the bandwidth on $D^{\parallel}/J$ in detail, we focus on the energy at the $\Gamma$ point of the degenerate modes 1 and 2.
Figure~\ref{fig:J_estimation}(a) shows the energy value as a function of $D^{\parallel}/J$.
It can be seen that the bandwidth increases linearly with $D^{\parallel}/J$.
Experimental results for Cs$_2$Cu$_3$SnF$_{12}$ show that the two high-energy modes are almost degenerate at the $\Gamma$ point, and their energies are about 10 meV.\cite{Saito2022,Ono2014}
By fixing $J=20.7$meV, which has been determined from magnetic susceptibility measurements,
and searching the value of $D^{\parallel}$ such that the calculated energy of the high-energy modes at the $\Gamma$ point is 10 meV, we obtain $D^{\parallel}/J =0.12$ (See Fig.~\ref{fig:J_estimation}(b)).

\begin{figure}[t]
\centering
\includegraphics[clip,width=5.3cm]{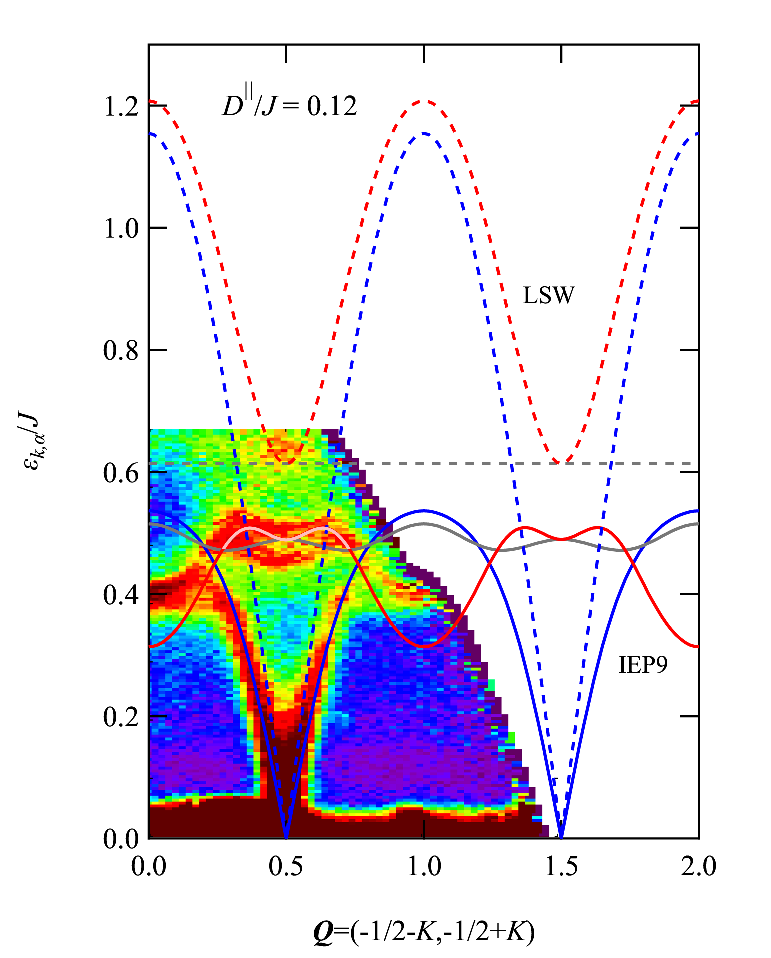}
\caption{(Color online) Comparison of magnon dispersions for $D^{\parallel}/J=0.12$ calculated by the LSW (dashed line) and the IEP with ninth order (solid line). 
The color map represents the result of the neutron scattering experiment with incident neutron energies of $E_{\rm i}=17.9$ meV.\cite{Saito2022}
}
\label{fig:lsw_vs_se}
\end{figure}

Figure~\ref{fig:lsw_vs_se} shows a comparison of the magnon dispersions between IEP and LSW, where $D^{\parallel}/J=0.12$ is chosen.
It can be seen that while the bandwidth of LSW is $1.2J$, the bandwidth of IEP is $0.5J$, and thus, the LSW magnon band is compressed by about 60\%.
In the case of a triangular lattice, the bandwidth of LSW is $1.6J$, while the bandwidth of IEP is $0.95J$, which indicates about 40\% compression.\cite{Zheng2006}
Such compression of the LSW magnon band may be a universal phenomenon in $S=1/2$ two-dimensional frustrated systems with magnetic ground states.
If so, the reason behind this may be that the spinon continuum is pushing the magnon band from above.\cite{Zhang2020}

The color map in Fig~\ref{fig:lsw_vs_se} is the experimental dynamical structure factor.
The incident neutron energy, $E_{\rm i}=17.9$ meV, is lower than that shown in Fig.~\ref{fig:dminteraction_direction}(b), and we can see more detailed structures of magnon dispersions.
It can be seen that the highest magnon branch takes maximum energy at points ($K\simeq 0.5\pm 0.15$) that is slightly away from the $\Gamma$ point ($K=0.5$).
It is possible to interpret it in light of our calculation as follows.
The LSW energy of mode 2 increases for $\Gamma\rightarrow {\rm M}$, but the renormalization factor $r_{k,2}$ decreases. 
Due to this competition, the energy of mode 2 reaches its maximum at points shifted from the $\Gamma$ point.
Near the M point, only one branch is observed in the experiment.
Our calculation suggests that this branch is mode 2, not mode 1, although the calculated energy of mode 2 is somewhat lower than that in the dynamic structure factor.
The discrepancy in energy values should be resolved by making the Hamiltonian more realistic.

In summary, we have used the IEP to investigate how spin-wave interactions modify the LSW predictions of magnon dispersions.
We have successfully calculated the renormalization factor $r_{k,\alpha}$ by the IEP.
As a result, in the kagome lattice system, in addition to the renormalization of the interaction parameter pointed out by Ono {\it et al.},\cite{Ono2014} we have found that the high-energy part around the M point undergoes strong deformation.

\section*{Acknowledgment}

\begin{acknowledgment}

The authors would like to thank the Supercomputer Center, Institute for Solid State Physics, University of Tokyo for the use of their facilities.

\end{acknowledgment}

\end{document}